\documentstyle[psfig,aps]{revtex}

\newcommand{\ga}{\alpha}
\newcommand{\gb}{\beta}
\newcommand{\gc}{\gamma}
\newcommand{\gd}{\delta}
\newcommand{\gk}{\kappa}
\newcommand{\gve}{\varepsilon}
\newcommand{\gl}{\lambda}
\newcommand{\gs}{\sigma}

\newcommand{\go}{\omega}

\newcommand{\gO}{\Omega}

\begin{document}
\title{ Deuteron life-time in hot and dense nuclear matter near equilibrium}
\author{M. Beyer and G. R\"opke
} 
\address{
FB Physik,
 Universit\"at Rostock, Universit\"atsplatz 1, 
18051 Rostock, Germany}

\maketitle

\begin{abstract}
  We consider deuteron formation in hot and dense nuclear matter close to
  equilibrium and evaluate the life-time of the deuteron fluctuations
  within the linear response theory.  To this end we derive a
  generalized linear Boltzmann equation where the collision integral
  is related to equilibrium correlation functions.  In this framework
  we then utilize finite temperature Green functions to evaluate the
  collision integrals. The elementary reaction cross section is
  evaluated within the Faddeev approach that is suitably modified to
  reflect the properties of the surrounding hot and dense matter.
\end{abstract}

\section{Introduction}
The complicated dynamics of heavy-ion collisions provides a great
challenge for many-particle theory.  In particular at intermediate
energies where the elementary ingredients are rather well known -- in
terms of constituents and their respective interactions -- the main
problem arises from a sufficient description of the many-particle
aspect. To provide single-particle distribution functions such
reactions can be simulated on the basis of kinetic equations as, e.g.,
supplied by the Boltzmann-Uhlenbeck-Uhling (BUU) approach (see, e.g.,
Refs.~\cite{sto86,ber88,cas90,bau92,bau93,alm95}).

However, the formation of light clusters such as deuterons, helium,
$\alpha$-particles, etc., is an important phenomenon of heavy-ion
collisions at intermediate energies, see, e.g., Ref.~\cite{nag81}.
Empirical evidence, including recent experimental data on cluster
formation~\cite{GSI93,MSU95}, indicate that a large fraction of
deuterons can be formed in heavy-ion collisions of energies below
$E/A\le 200$ MeV. Also, during the expansion of the system the density
can drop below the Mott-density of deuteron
dissociation~\cite{bal95,sch90,alm90}.

The description of the formation of such bound states (clusters)
during the expansion of hot and dense matter is not as well elaborated
as the single-particle distribution. The main obstacle is that the
formation of bound states requires the notion of few-body reactions
within the medium. Even the simplest case, i.e., the abundances of
deuterons that are determined by the deuteron formation via
$NNN\rightarrow dN$ ($N$ nucleon, $d$ deuteron) and break-up,
$dN\rightarrow NNN$, reactions, requires a proper treatment of the
effective three-body problem.  Previous studies of the kinetics of
deuteron production have utilized the impulse approximation to
calculate the reaction cross section at energies above 200
MeV~\cite{dan91}. For lower energies, viz. $E/A \le 200$ MeV, the
impulse approximation fails badly and a full three-body treatment of
the scattering problem is necessary~\cite{bey96}. Furthermore, a
consistent treatment of cluster formation in expanding hot and dense
matter requires the inclusion of medium effects into the respective
elementary reaction cross sections as has been done in the nucleon
nucleon (NN)-case and proven to be substantial in BUU-simulations of
heavy-ion reactions~\cite{alm95}.  Therefore we present an exact
treatment of the three-body problem including medium modifications in
mean-field approximation.

The cluster formation during the expansion of hot and dense matter is
driven by the collision term in the generalized linear Boltzmann
equation (see, e.g., Ref.s~\cite{roe88,zub96}).  Here we consider the
fluctuations of the deuteron distributions in hot and dense nuclear
matter in a near-equilibrium situation. One important question in this
context is the time scale of formation and disintegration processes
that govern the evolution to chemical equilibrium.  Within linear
response theory we relate the reaction rates to the equilibrium
correlation functions. To calculate the response coefficients it is
then possible to apply the method of finite temperature Green
functions~\cite{kad62,kra86,kap89}.

The essentials of the three-body problem for the isolated system are
well known, see, e.g., Ref.~\cite{glo88}.  In the following we utilize the
AGS-formalism~\cite{alt67} suitably modified to treat the three-body
problem within nuclear matter. To derive the proper AGS-type equations
we use the self-consistent random phase approximation~\cite{sch73}
extended to finite temperatures.  For a numerical solution we rely on
a separable representation of the NN-potential.  This choice
simplifies the problem considerably.  A systematic investigation of
separable parameterizations of ``realistic'' potentials has been
pursued, e.g., in Refs.~\cite{ple95}.  We note
that solutions of the three-body problem using ``realistic''
NN-potentials have been achieved, e.g., by the Bochum
group~\cite{glo90}, and the Bonn group in the framework of the
$W$-matrix approach~\cite{san90}.

In the following section we present the formalism to treat cluster
formation in a linear approximation of the generalized Boltzmann
equation. In Sec. III we introduce the finite temperature Green
function and derive a Faddeev-type equation that includes medium
modifications due to Pauli-blocking and energy-shifts. We relate the
``in-medium'' cross section to the collision term in the Boltzmann
equation.  Our numerical results are presented in Sec. IV and we
summarize and conclude in Sec. V.

\section{Quantum kinetics and bound state formation}

The Hamiltonian of the Fermi system in question is given in terms of
creation and annihilation operators,
\begin{equation}
H=\sum_{11'}H_0(1,1') a^\dagger_{1}a_{1'}
+\frac{1}{2}\sum_{121'2'}V_2(12,1'2') 
a^\dagger_{1}a^\dagger_{2}a_{2'}a_{1'}, 
\end{equation}
where $ a^\dagger$, $a$ 
satisfy the well known commutation relations. The indices $1,2,\dots$
collectively denote the quantum numbers (e.g.,  momentum, spin,
isospin,\dots) of the particles $1,2,\dots$  The observed physical
quantities will be expressed in terms of reduced $n$-particle
occupation matrices (see, e.g., Refs.~\cite{zub96,mor95}),
\begin{equation}
f_n(1\dots n,1'\dots n';t)= 
\langle a^\dagger_{n'}\dots a^\dagger_{1'} a_{1}\dots a_{n}\rangle^t 
\equiv {\rm Tr}\{ \rho(t)\;  a^\dagger_{n'}\dots a^\dagger_{1'} 
a_{1}\dots a_{n}\},
\label{eqn:dist}
\end{equation}
where $\rho(t)$ denotes the density matrix of the many-particle
system. In case of equilibrium ($\rho(t)=\rho_0$) we use the notation
$f^0_n(\dots)=\langle\dots\rangle_0$.  

The reduced density matrices given above are particularly suited in
the framework of a cluster decomposition~\cite{zub96}.  If clusters
are treated in mean-field approximation we may introduce cluster wave
functions $\varphi_{\nu}$ and introduce bosonic (two-particle)
operators that are given through
\begin{equation}
b_{\nu} = \sum_{12}\;  a_{1}a_{2}\;
\langle \varphi_{\nu}|12\rangle,
\label{eqn:fock}
\end{equation}
and the h.c. $b^\dagger$. For the two-particle system of interest
here $\varphi_{\nu}$ is given by the solution of the respective two
particle Bethe-Salpeter equation with the eigenvalues $E_\nu$ (of
bound or scattering states)~\cite{sch90}. 
This way it is possible to write, e.g.,  $f_2$ in
a cluster representation, viz.
\begin{equation}
f_2(\nu,\nu';t)= \langle b_{\nu'}^\dagger b_{\nu}\rangle^t.
\end{equation}

For nuclear matter the conditions in the final stage of a heavy-ion
collision may be such that formation of bound states is possible. This
is indeed the case, when the density of the system is below the Mott
density, e.g., of the deuterons, and formation will
occur~\cite{bal95,sch90,alm90}. Following a general density matrix
approach as given in Ref.~\cite{roe88A} the time evolution of the
distribution functions for nucleons $f_1(t)\equiv f_{N}(p,t)$ with
momentum $p$ and deuterons $f_2(t)\equiv f_{d}(P,t)$ with momentum $P$
reads for homogeneous matter
\begin{eqnarray}
\partial_t f_{N}(p,t) &= &{\rm Tr}\left\{\rho(t) 
  i[H,n_{Np}]\right\}
= - {\cal D}_{N}(p,t) + {\cal I}_{N}(p,t),\\
\partial_t f_{d}(P,t) &= &{\rm Tr}\left\{\rho(t)
   i[H,n_{dP}]\right\}
= - {\cal D}_{d}(P,t) + {\cal I}_{d}(P,t),
\end{eqnarray} 
where $n_{Np}=a_{Np}^\dagger a_{Np}$ and $n_{dP}=b_{dP}^\dagger
b_{dP}$. The Vlasov terms ${\cal D}(t)$ describe the reversible time
evolution and are related to time dependent Hartree-Fock calculations
as, e.g., explained in Ref.~\cite{roe88A}.

The collision terms ${\cal I}(t)={\cal I}^{\rm E}(t)+{\cal I}^{\rm
  R}(t)$ correspond to the irreversible behavior and describe elastic
scattering (E) and inelastic (reaction) processes (R), respectively,
between the constituents of the system.  The elastic processes do not
change the internal quantum numbers of the particles. They determine
the time scales for thermal relaxation.  However, the inelastic
processes that are related to excitation as well as to bound state
formation and disintegration change the abundances of the components
characterized by the internal quantum numbers and determine the time
scale of chemical equilibration. The reaction relevant for the energy
domain considered are due to photodisintegration $ I_{\gc NN,d}(p,t)$
and nucleon deuteron break-up $I_{NNN,Nd}(p,t)$ (and the
reversed ones), i.e.
\begin{eqnarray}
{\cal I}_{N}^{\rm R}(p,t)&=&  I_{\gc NN,d}(p,t) + I_{NNN,Nd}(p,t)
+\dots
\label{eqn:IN}\\ 
{\cal I}_{d}^{\rm R}(P,t)&=&I_{d,\gc NN}(P,t)  
+ I_{dN,NNN}(P,t)+\dots\label{eqn:Id}
\end{eqnarray}
Further reaction channels represented by the dots are given in
Ref.~\cite{roe88}. Presently, we consider the three-particle
processes.

The collision integral that involve three-particle processes have been
given in Born approximation~\cite{roe88A} or evaluated in impulse
approximation~\cite{dan91}. In both cases the influence of the
surrounding medium on the elementary cross section that enter into the
collision integrals has been neglected.  This might not be sufficient
for intermediate energy heavy-ion reactions as has been shown, e.g.,
for the NN collision rate in a BUU calculation of La on
La~\cite{alm95}. The approach presented here naturally respects medium
modification in the break-up cross sections.  Furthermore, in view of
the rather moderate energies reached, the effective three-body problem
arising in this context is treated exactly in terms of properly
generalized Faddeev-type equations.

To be more specific we consider the situation where the collision rate
is sufficiently high compared to the reaction rate, so that each
components are close to their thermal equilibrium distributions. The
small deviations of the chemical composition from equilibrium are
then treated within the linear response theory. The time scale of the
relaxation to chemical equilibrium is set by the reaction processes that will
be considered in the following. In this case 
\begin{eqnarray}
\partial_t f_{N}^{\rm R}(p,t) &= &I_{N}^{\rm R}(p,t),\label{eqn:RN}\\
\partial_t f_{d}^{\rm R}(P,t) &= &I_{d}^{\rm R}(P,t),\label{eqn:RD}
\end{eqnarray} 
where we have introduced $I_{N}^{\rm R}(p,t)= I_{NNN,Nd}(p,t)$ and
$I_{d}^{\rm R}(P,t)= I_{dN,NNN}(P,t)$ for brevity. The collision terms
on the rhs. of Eqs.~(\ref{eqn:RN}) and (\ref{eqn:RD}) each contain
gain and loss terms due to deuteron break-up or formation reactions.
Collisions of higher clusters (e.g., $dd$) that require a suitable
treatment of the effective four-body problem are left for further
investigations. To evaluate the integral $I_{d}^{\rm R}(P,t)$ we use
linear response theory (see appendix  A),
\begin{equation}
I_{d}^{\rm R}(P,t) = - \sum_{P'P''}\langle \dot n_{dP};\dot n_{dP'}\rangle
(n_{dP'};n_{dP''})^{-1}\;\gd f_{d}(P'',t),
\label{eqn:fluk}
\end{equation}
where $\gd f_{d}(P,t) = f_{d}(P,t)- f_{d}^0(P)$ denotes the
fluctuations from the equilibrium distribution, and $n_{dP}=\langle
b^\dagger_{dP} b_{dP}\rangle$. The Kubo scalar
product $(A;B)$ that appears in Eq.~(\ref{eqn:fluk}) is given in
Eq.~(\ref{eqn:kubo}) and its Laplace transform, i.e.  the correlation
functions $\langle A(\eta\rightarrow 0^+);B\rangle$, in
Eq.~(\ref{eqn:laplace}). Following standard many-body techniques (see,
e.g., Refs.~\cite{roe88,kad62,fet71}) the correlation function is
evaluated using Green functions,
\begin{equation}
\langle A(\eta);B\rangle=
-\frac{1}{\gb} \int \frac{d\go}{2\pi}\;
\frac{1}{\eta+i\go}\;\frac{1}{\go}
\;\left[G_{AB}(\go+i0^+)-G_{AB}(\go-i0^+)\right],
\label{eqn:Gcorr}
\end{equation}
where $G_{AB}(z)$ is the analytic continuation of the Matsubara
Green function $G_{AB}(z_\mu)$ that will be discussed in the next
section.

For homogeneous matter where $\langle \dot n_{dP};\dot
n_{dP'}\rangle$ and $(n_{dP'};n_{dP''})$ are diagonal in momenta $P$, 
the response equation (\ref{eqn:RD}) is given by
\begin{equation}
\partial_t\,f_{d}^{\rm R}(P,t) =  
\langle \dot n_{dP}(\eta\rightarrow 0^+);\dot n_{dP}\rangle
(n_{dP};n_{dP})^{-1}\;\gd f_{d}^{\rm R}(P,t) 
\equiv \frac{1}{\tau_{dP}}\;\gd f_{d}^{\rm R}(P,t).
\label{eqn:time}
\end{equation}
The limit $\eta\rightarrow 0^+$ implied through Eq.~(\ref{eqn:time})
has to be taken after the thermodynamic limit.  Here we have
introduced the momentum dependent life-time (formation-time) of the
deuteron fluctuations $\tau_{dP}$ in the surrounding medium, which is
of central interest. Note that the disintegration (formation) of
deuterons requires the explicit treatment of three-particle equations
in nuclear matter, which will be derived in the following section.

\section{Finite temperature Green function and three-body equations}
\label{sec:green}

In order to evaluate Eq.~(\ref{eqn:time}) by use of
Eq.~(\ref{eqn:Gcorr}) we need to define the finite temperature Green
function. To consider $n$ particles embedded in a medium the
$n$-particle Green function ${\cal G}_n(1\dots n,1'\dots n')$ for
equilibrium is defined by
\begin{equation}
{\cal G}^{t-t'}_n(1\dots n,1'\dots n') = -i\;
\langle T\;A_n(t)A^\dagger_{n'}(t)\rangle_0,
\label{eqn:def}
\end{equation}
where $T$ implies Wick time ordering. The operators $A_n(t)$ are
$n$-particle Operators, i.e. $A_n(t)=a_1(t_1)\dots
a_n(t_n)|_{t_1=\dots=t_n=t}$ taken at equal times, and in the
Heisenberg picture
\begin{equation}
A(t)=\exp(iHt)\;A\;\exp(-iHt).
\end{equation}
The Green functions given in Eq.~(\ref{eqn:def}) satisfy a hierarchy
of equations, given, e.g., in Refs.~\cite{kad62,kra86,kap89}. To arrive
at equations that are solvable in practice for the $n$-particle
problem, one has to truncate the hierarchy, which is usually done by
introducing suitable approximations for the $(n+1)$st-particle Green
function. For the three-particle problem this has been done, e.g., in
Ref.~\cite{fis70} in special cases. Within the self-consistent random
phase approximation  it is possible to arrive at equations that
are already decoupled. This methods has been used for zero
temperatures in Ref.~\cite{sch73} and will be extended to finite
temperatures here.  To simplify the notation we use a matrix form
in the following
\begin{equation}
{\cal G}^{t-t'}_n =
\left({\cal G}^{t-t'}_n(1\dots n,1'\dots n')\;\right).
\end{equation}
The time evolution of the $n$-particle Green function is governed
by a Dyson equation~\cite{sch73}
\begin{equation}
i\partial_t {\cal G}^{t-t'}_{n}
=\gd(t-t') {\cal N}^t_{n} 
+ \int d\bar t\; {\cal M}^{t-\bar t}_{n}
\;{\cal G}^{\bar t-t'}_{n}.
\label{eqn:greenG}
\end{equation}
The mass matrix $ {\cal M}^{t-\bar t}_{n}$ introduced in the
above equation is given by
\begin{equation}
 {\cal M}^{t-\bar t}_{n}
= \gd(t-\bar t) {\cal M}^{t}_{n,0}+
 {\cal M}^{t-\bar t}_{n,irr.}
\end{equation}
with
\begin{eqnarray}
\left({\cal M}^{t}_{n,0}{\cal N}_{n}\right)
(1\dots n,1'\dots n')
&=&{\rm Tr}\{\rho_0 \left[[A_n,H],A^\dagger_{n'}\right]_\pm\}
\label{eqn:mean}\\
\left({\cal M}^{t-\bar t}_{n,irr.}{\cal N}_{n'}\right)
(1\dots n,1'\dots n')
&=&-i{\rm Tr}\{\rho_0\;T [A_n,H]_t[H,A^\dagger_{n'}]_{\bar t}]\}_{irr.},
\end{eqnarray}
where the index $irr.$ indicates that all reducible parts should be
omitted, where the index $\pm$ refers to odd ($+$) or even ($-$)
number$n$  of fermions. The first term refers to an $n$-body cluster mean-field
contribution~\cite{kra86,sch73} whereas the second term is of
dynamical origin and contains retardation. Up to the correlations of
interest instantaneous and dynamical contributions are separated.  The
normalization is given by
\begin{equation}
{\cal N}_n(1\dots n,1'\dots n') = {\rm Tr}\{\rho_0\; 
[A_n,A_{n'}^\dagger]_\pm\}.
\end{equation}
In cluster mean-field approximation the term $ {\cal M}_{n,irr.}$ will
be neglected. 
In the Matsubara-Fourier representation the Green function is given by
\begin{equation}
{\cal G}^{t-t'}_n = \frac{1}{-i\gb}
\sum_{\mu}\;e^{iz_\mu(t-t')}\;G_n(z_\mu).
\end{equation}
For a fermionic system considered here $z_\mu$ is a fermionic or
bosonic Matsubara frequency, depending on whether $n$ is odd or even,
resp. to preserve the Kubo-Martin-Schwinger boundary
condition~\cite{kad62,kra86,kap89}.

Taking in Eq.~(\ref{eqn:greenG}) for $A_3=a_1a_2a_3$ and evaluating
Eq.~(\ref{eqn:mean}) in the independent particle approximation leads
to the following Bethe-Salpeter equation for the three-particle Green
function at finite temperatures and densities,
\begin{equation}
G_3(z_\mu) = G_3^{(0)}(z_\mu) + R_3^{(0)}(z_\mu)
  \tilde V_3 G_3(z_\mu),
\label{eqn:G3}
\end{equation}
which is the central input to derive Faddeev type equations in a
medium. The notation will be explained in the following. The proper
symmetrization is treated separately.  The Green function of the
noninteracting system is
\begin{equation}
G_3^{(0)}(z_\mu)=N_3R_3^{(0)}(z_\mu),
\label{eqn:G0}
\end{equation}
where
\begin{eqnarray}
 R_3^{(0)}(123,1'2'3';z_\mu) &= &\frac{\gd_{11'}\gd_{22'}\gd_{33'}}
{z_\mu - \gve_1 - \gve_2 - \gve_3},\\
N_3(123,1'2'3') &= &\gd_{11'}\gd_{22'}\gd_{33'}
(f_1f_2f_3 + \bar f_1 \bar f_2 \bar f_3)\label{eqn:FPauli}\\
&=&\gd_{11'}\gd_{22'}\gd_{33'}
(1-f_i-f_j)(1-f_k+g(\gve_i+\gve_j)).
\label{eqn:Falt}
\end{eqnarray}
Note that Eq.~(\ref{eqn:Falt}) is identical to Eq.~(\ref{eqn:FPauli})
for all permutations of $ijk=123$.  We use $\bar f = 1-f$, and the
Fermi one-particle function $f(\gve_1)= (\exp[\gb(\gve_1-\mu)]+1)^{-1}$ and the
Bose function $g(\omega)= (\exp[\gb(\omega-2\mu)]-1)^{-1}$ for the two
fermion system.
Here $\gb$ is the inverse temperature of the system
and $\mu$ is the chemical potential. In mean field approximation the
single quasi-particle energy $\gve_1$ is given by
\begin{eqnarray}
\gve_1 &= &\frac{k^2_1}{2m_1} + \Sigma^{HF}(1), \\
\Sigma^{HF}(1) &=& \sum_{2}
\left[ V_2(12,12)-  V_2(12,21) \right]\,  f_2.
\end{eqnarray}
Note that $[N_3,R_3^{(0)}]=0$.  The interaction kernel in $ \tilde V_3$
in Eq.~(\ref{eqn:G3}) is given by
\begin{eqnarray}
  \tilde V_3(123,1'2'3') &=&\sum_{k=1}^3 \tilde V^{(k)}_3(123,1'2'3'),
\label{eqn:Vchn}\\
\tilde  V^{(k)}_3(123,1'2'3')&=& (1-f_i-f_j) V_2(ij,i'j')\gd_{kk'},
\label{eqn:V3}
\end{eqnarray}
with $ijk=123$ cyclic, and $\tilde V_3\neq\tilde
V_3^\dagger$. If we introduce a potential $V_3 = N_3^{-1}\tilde V_3$ we
may instead of Eq.~(\ref{eqn:G3}) write
\begin{equation}
G_3(z_\mu) = G_3^{(0)}(z_\mu) + G_3^{(0)}(z_\mu) V_3 G_3(z_\mu),
\label{eqn:G3neu}
\end{equation}
which looks formally as the equation for the isolated case~\cite{glo88}.
Using Eq.~(\ref{eqn:Falt}) we may write the potential $V_3=\sum_k
V_3^{(k)}$ in terms of (e.g., $k=1$)
\begin{equation}
V^{(1)}_3(123,1'2'3')= (1-f_1+g(\gve_2+\gve_3))^{-1}
V_2(23,2'3')\gd_{11'},
\label{eqn:V3neu}
\end{equation}
 
In Eq.~(\ref{eqn:Vchn}) we have already introduced the channel
notation that is convenient to treat systems with more than two
particles~\cite{glo88}. 

If the pair and the odd particle are uncorrelated in channel $\gc$,
we may define a channel Green function $G_3^{(\gc)}(z_\mu)$.  In this
case only the interaction within the pair of channel $\gc$ is taken
into account, viz.
\begin{equation}
G_3^{(\gc)}(z_\mu) = \frac{1}{-i\gb} \sum_\gl\;
iG_2(\go_\gl)\;G_1(z_\mu-\go_\gl). 
\label{eqn:Gchanneldef}
\end{equation}
The summation is done over the Bosonic Matsubara frequencies
$\go_\gl$, $\gl$ even, $\go_\gl=\pi\gl/(-i\gb) +2\mu$.  The equation
for the channel Green function is derived in the same way as for the
total three particle Green function given in Eqs.~(\ref{eqn:G3}) and
(\ref{eqn:G3neu}). The result is
\begin{equation}
G_3^{(\gc)}(z_\mu) = G_3^{(0)}(z_\mu) + G_3^{(0)}(z_\mu)
  V^{(\gc)}_3 G_3^{(\gc)}(z_\mu).
\label{eqn:Gchannel}
\end{equation}
Introducing the notation $\bar V_3^{(\gc)} = V_3 - V_3^{(\gc)}$ we arrive
at the following equation for $G_3(z_\mu)$ expressed through the
channel Green functions $G_3^{(\gc)}(z_\mu)$, i.e.
\begin{equation}
G_3(z_\mu) = G_3^{(\gc)}(z_\mu)
+G_3^{(\gc)}(z_\mu) \bar V_3^{(\gc)} G_3(z_\mu).
\label{eqn:G3Ggam}
\end{equation}

Now we have set the necessary equations, i.e. Eqs.~(\ref{eqn:G3neu}),
(\ref{eqn:Gchannel}) and (\ref{eqn:G3Ggam}) to define a channel
transition operator $U_{\ga\gb}$ fir finite temperature,
\begin{equation}
G_3(z_\mu)=\gd_{\ga\gb} G_3^{(\ga)}(z_\mu)+ 
G_3^{(\ga)}(z_\mu)U_{\ga\gb}(z_\mu)G_3^{(\gb)}(z_\mu).
\end{equation}
In the zero density limit, this definition coincides with the usual
definition of the transition operator with the correct reduction
formula to calculate cross sections~\cite{san72}. Inserting this
definition into Eq.~(\ref{eqn:G3Ggam}) and using
Eq.~(\ref{eqn:Gchannel}) leads to an equation for the
transition operator in medium, viz.
\begin{equation}
U_{\ga\gb} = (1-\gd_{\ga\gb}) G_3^{(0)-1}+ 
\sum_{\gc\neq \ga}   V_3^{(\gc)} G_3^{(\gc)}U_{\gc\gb}.
\end{equation}
This is the AGS-type (or Faddeev-type) equation valid to treat
three-particle correlations at finite temperatures and densities in
mean-field approximation.  Although this equation looks formally equal
to that for the isolated system, we emphasize that $V_3^{(\gc)}$ as
well as $G_3^{(\gc)}$, and hence $U_{\gc\gb}$ is different from the
isolated system due to the finite temperature and density of the
surrounding matter, and therefore contains Pauli factors due to phase
space occupation and self-energy shifts. This becomes transparent, if
the definitions of the quantities appearing in this equation are
inserted. Before doing that we define a transition channel operator $
T^{(\gc)}_3$,
\begin{equation}
G_3^{(\gc)}= G_3^{(0)}+ 
G_3^{(0)}  T^{(\gc)}_3 G_3^{(0)},
\end{equation}
inserting this equation into Eq.~(\ref{eqn:Gchannel})
leads to
\begin{equation}
  T^{(\gc)}_3 =   V_3^{(\gc)} + 
G_3^{(0)}  V_3^{(\gc)}   T_3^{(\gc)},
\label{eqn:Tchannel}
\end{equation}
and to $V_3^{(\gc)} G_3^{(\gc)} = T_3^{(\gc)} G_3^{(0)}$.
This way it is possible to write a second, more useful version of the
AGS-type equations
\begin{equation}
U_{\ga\gb} = (1-\gd_{\ga\gb}) (N_3R_3^{(0)})^{-1}+ 
\sum_{\gc\neq \ga}   T_3^{(\gc)} N_3 R_3^{(0)}U_{\gc\gb}.
\label{eqn:AGS}
\end{equation}
Here we have written the Pauli factors occuring  due to the
surrounding matter explicitly. Note, that through Eq.~(\ref{eqn:Tchannel})
$T_3^{(\gc)}$ is also medium dependent.  To compare with our previous
  result~\cite{bey96}, we repeat the expressions for the low density
  case. In this case, we may assume $N_3>0$, and therefore write an
  equation for $U^*_{\gc\gb}=N^{1/2}_3U_{\gc\gb}N^{1/2}_3$,
\begin{equation}
U^{*}_{\ga\gb}
= (1-\gd_{\ga\gb}) R_3^{(0)-1}+ 
\sum_{\gc\neq \ga} T_3^{*(\gc)} R_3^{(0)}U^*_{\gc\gb}.
\label{eqn:AGSlow}
\end{equation}
The equation for the transition channel operator $T_3^{*(\gc)}
=N_3^{1/2} T_3^{(\gc)} N_3^{1/2}$ (and so for $V_3$) is then
\begin{equation}
T_3^{*(\gc)}= V_3^{*(\gc)}+V_3^{*(\gc)}R_3^{(0)}T_3^{*(\gc)}.
\end{equation}
Inserting all definitions
the explicit form of the effective potential arising in this equation
reads
\begin{eqnarray}
V_3^{*(3)}(123,1'2'3')
&=&(1-f_1-f_2)^{1/2}(1-f_3+g(\gve_1+\gve_2))^{-1/2} \nonumber\\
&&\times V_2(12,1'2')\gd_{33'}
(1-f_3+g(\gve_{1'}+\gve_{2'}))^{1/2}(1-f_{1'}-f_{2'})^{1/2}\\
&\simeq&(1-f_1-f_2)^{1/2} V_2(12,1'2')(1-f_{1'}-f_{2'})^{1/2},
\label{eqn:hemitian}
\end{eqnarray}
where  Eq.~(\ref{eqn:hemitian}) holds for $f^2\ll f$. Utilizing
this approximation Eq.~(\ref{eqn:AGSlow}) has been solved numerically
using a separable ansatz for the strong nucleon-nucleon
potential~\cite{bey96}.

We are now in the position to evaluate the correlation function $
\langle \dot n(\eta\rightarrow 0^+);\dot n\rangle$ according to
Eq.~(\ref{eqn:Gcorr}). The Green function needed in
Eq.~(\ref{eqn:Gcorr}) is given by (see appendix B)
\begin{eqnarray}
G^{(\gc)}_{\dot n_{dP} \dot n_{dP}} (\gO_\mu) 
&=& \frac{4}{-i\gb}\sum_\gl
\;\mbox{Tr}\left\{U_{\gc 0}\;G^{(0)}_3(\gO_\mu+z_\gl)
\;U_{0\gc }\;G^{(\gc)}_{3,dP}(z_\gl)\right\}
\label{eqn:G6full}\\
&&\qquad\qquad+ (\gO_\nu\leftrightarrow -\gO_\nu)\nonumber.
\end{eqnarray}
To perform the Matsubara summation that is present in
Eq.~(\ref{eqn:G6full}), we now use the spectral representation of the
Green functions that have been given for the quasi-particle
approximation in Eqs.~(\ref{eqn:G0}) and (\ref{eqn:Gchanneldef}).  The 
resulting expression for $G_{\dot n_k \dot n_p} (\gO_\mu)$ may be cast
into the following form,
\begin{eqnarray}
G^{(\gc)}_{\dot n_{d} \dot n_{dP}} (\gO_\mu) 
&=& 4 i\sum_{123c}\frac{
 \langle 123|U_{\gc 0}|c\varphi^{(\gc)}_{dP}\rangle
\langle c\varphi^{(\gc)}_{dP}|U_{0\gc}|123\rangle}
{\gO_\mu + (E^{(\gc)}_{dP}+\gve_c) - E_0}\nonumber\\
&&\qquad\times[\bar f_{1}\bar f_{2}\bar f_{3}f_{c}g(E^{(\gc)}_{dP})
-f_{1}f_{2}f_{3}\bar f_{c}(1+g(E^{(\gc)}_{dP}))]\label{eqn:G6fine}\\
&&\qquad+ (\gO_\nu\leftrightarrow -\gO_\nu)\nonumber,
\end{eqnarray}
with $E_0=\gve_{1}+\gve_{2}+\gve_{3}$, and $c\in\{1'2'3'\}$ depending
on the channel $\gc\in\{123\}$, resp. The terms in brackets of
Eq.~(\ref{eqn:G6fine}) are usually referred to as Pauli factors of the
gain and loss terms. Since presently we are interested in the time
scale of fluctuations, we may consider, e.g., the loss term  that
is given by
\begin{equation}
\langle \gd \dot n^{(\gc)}_{dP}
;\gd\dot n^{(\gc)}_{dP}\rangle
= 4 \sum_{123c} 
|\langle 123|U_{0\gc}|\varphi^{(\gc)}_{dP},c\rangle|^2
 \bar f_{1}\bar f_{2}\bar f_{3}f_{c}g(E^{(\gc)}_{dP})
\;2\pi\gd(E^{(\gc)}_{dP}+\gve_c - E_0).
\label{eqn:corrfunc}
\end{equation}
For identical particles that are considered here proper symmetrization
has to be taken into account. To this end we connect the result given
in Eq.~(\ref{eqn:corrfunc}) with the transition matrix $T_0$, that is
evaluated between properly symmetrized and normalized states $\phi_0$,
$\phi_{d}$ and again satisfies the equivalent three-body equation, see
Refs.~\cite{glo88} for the isolated case. It is given by
\begin{equation}
 \langle \phi_0|T_{0}|\phi_{d}\rangle\equiv
\sqrt{2} \sum_\gc
\langle 123|U_{0\gc}|\varphi^{(\gc)}_{dP},c\rangle.
\end{equation}
To be more specific we also separate spin and momentum degrees of
freedom and introduce amplitudes
\begin{equation}
\langle m_1m_2m_3|{\cal M}_0({\bf k}_1{\bf k}_2{\bf k}_3,
{\bf k}_N{\bf P};E) |m_jm_2\rangle 
\equiv \langle \phi_0|T_{0}|\phi_{d}\rangle.
\end{equation}
Using this amplitudes the life-time  may be written in the
following way
\begin{equation}
\tau^{-1}_d 
= \frac{4}{3!} \int d^3 k_N\int d^3k_1 d^3k_2 d^3k_3\;
{\rm Tr}({\cal M}_0\rho_i{\cal M}_0^\dagger)
\; \bar f_{1}\bar f_{2}\bar f_{3}f_{\gve}
\;2\pi\gd(E - E_0).
\label{eqn:lifetime}
\end{equation}
The total energy is given by $E=P^2/2m_d + \gve + E_d$ and
$\gve=k_N^2/2m$ is the energy of the odd nucleon.  The factor $1/3!$
prevents overcounting due to the six possible ways of arranging the
identical particles among the three  momenta, the trace is
over spin projections only and $\rho_i$ is the initial spin density
matrix.

It is instructive to discuss several ways to recover the Born
approximation and impulse approximations that have been used
previously~\cite{dan91,roe87}.
From Eq.~(\ref{eqn:AGS}) the lowest order iteration for the (on-shell)
break-up amplitude $U_{\gc0}$ is
\begin{equation}
U_{\gc 0} \simeq \sum_{\gd\neq\gc} T_3^{(\gd)} 
\simeq  \sum_{\gd\neq\gc} V_3^{(\gd)} = \bar V_3^{(\gc)}, 
\end{equation}
where the first term is referred to as impulse and the second as Born
approximation. Replacing  $U_{\gc 0}$ by $\bar V^{(\gc)}_3$ in
Eq.~(\ref{eqn:G6full}) leads to
\begin{eqnarray}
G^{(\gc){\rm Born}}_{\dot n_{dP} \dot n_{dP}} (\gO_\mu) 
&=& \frac{4}{-i\gb}\sum_\gl
\mbox{Tr}\left\{\;\bar V^{(\gc)}_{3}
\;G^{(0)}_3(\gO_\mu+z_\gl)
\;\bar V^{(\gc)}_{3}\;G^{(\gc)}_{3,dP}(z_\gl)\right\}
\label{eqn:G6Born}\\
&&\qquad+ (\gO_\nu\leftrightarrow -\gO_\nu)\nonumber,
\end{eqnarray}

A second possibility is to expand the Green functions into the
spectral representation. 
In Eq.~(\ref{eqn:G6full}) the spectral representation of the Green
functions leads to matrix elements $\langle \phi^{(\gc)}_{dP}|U_{\gc
  0}|\phi_0\rangle$.  Then we may use the on-energy-shell relation
\begin{equation}
\langle \phi_0|U_{0\gc}|\phi^{(\gc)}_{dP}\rangle=
\langle \phi_0|\bar V^{(\gc)}|\Psi^{(\gc)}_{dP}\rangle^{(+)}
\rightarrow
\langle \phi_0|\bar V^{(\gc)}|\phi^{(\gc)}_{dP}\rangle,
\end{equation}
which in turn after reinserting the spectral expansion leads again to
Eq.~(\ref{eqn:G6Born}).

\section{Results}

Although formally rather simple the integration over the 
momenta $k_i$ of Eq.~(\ref{eqn:lifetime}) is rather tedious, since momenta
dependencies appear also in the Fermi functions $\bar f$. In the low
density approximation we may as indicated at the end of Sec.
\ref{sec:green}, use the definition $U = N_3^{1/2} U^* N_3^{1/2}$, which
leads to ${\cal M}{\cal M}^{\dagger} = N_{3}^{-1}{\cal
  M}^*{\cal M}^{*\dagger} N_{3}^{-1}$, and
\begin{equation}
 N_{3}\simeq \bar f_{1}\bar f_{2}\bar f_{3}.
\end{equation}
Hence the term $ \bar f_{1}\bar f_{2}\bar f_{3}$ in
Eq.~(\ref{eqn:lifetime}) will be absorbed in the redefinition of
${\cal M}^*$.  The additional Fermi functions $( \bar f_{1}\bar
f_{2}\bar f_{3})^{-1}$ appearing in the $Nd$ channel
due to the replacement ${\cal M}_0\rightarrow {\cal M}_0^*$ will be
approximated in the following way (note $f^2\ll f$)
\begin{equation}
f_{\gve}/( \bar f_{1}\bar f_{2}\bar f_{3})
\simeq f_{\gve}(1+f_{1}+f_{2}+f_{3})\simeq f_{\gve}.
\end{equation}

We may now introduce the in-medium break-up cross section $\gs^*_0$ in
the center of mass system, which coincides with the usual in-vacuum
definition in the zero density limit~\cite{bey96}.
It is given by
\begin{equation}
  \gs^*_0(E) = \frac{(2\pi)^3}{|{\bf v}_d-{\bf v}_N|}
\;\frac{1}{3!}
\int d^3p' d^3q' \;{\rm Tr}({\cal M}^*_0\rho_i{\cal M}_0^{*\dagger})
\;2\pi\gd(E^*-E^*_0).
\label{eqn:cross}
\end{equation}
The cross section is evaluated in the center of mass system
introducing Jacobi coordinates ${\bf p}'$, ${\bf q}'$, and $|{\bf
  v}_d-{\bf v}_N|$ is the relative velocity of the incoming particles.
The center of mass scattering energy is $E^*=3q^2/4m^* + E^*_d$, where
we have used effective mass approximation for the nucleon self
energy~\cite{bey96}. 
Due to the medium effects the deuteron binding energy changes, which
is calculated consistently with the two-body input into the Faddeev
equation that lead to the amplitudes ${\cal M}^*_0$.  To evaluate the
life-time of the deuteron in medium we introduce the cross section
defined in Eq.~(\ref{eqn:cross}) into Eq.~(\ref{eqn:lifetime}). The
remaining integration is over the momentum ${\bf k}_N$ of the odd
nucleon.  The equation for the life-time of the deuteron is then given
by
\begin{equation}
\tau^{-1}_{dP} = \frac{4}{(2\pi)^3}
\;\int d^3k_N\; |{\bf v}_d-{\bf v}_N|\;\gs^*_0(E^*)\;f(\gve^*),
\label{eqn:liferes}
\end{equation} 
and $\gve^*=(3{\bf q}/2+{\bf P}/2)^2/2m^*$.

\begin{figure}[p]
  \leavevmode
\centering
  \psfig{figure=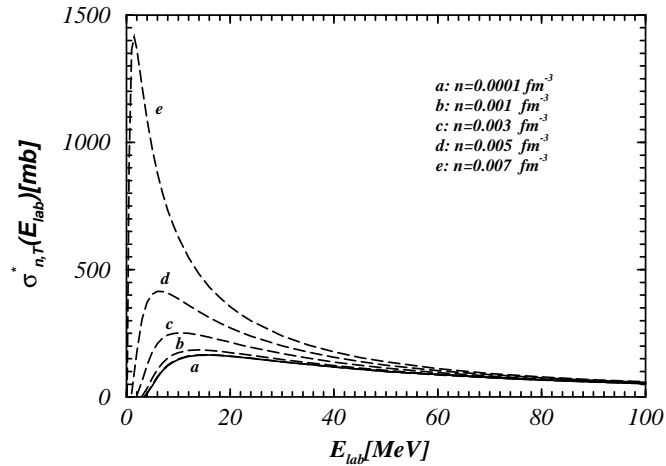,angle=270,width=0.6\textwidth}
\vspace*{1ex}
\caption{\label{fig:cross} Break-up cross section at temperature $T=10$
  MeV. Free cross section is shown as solid line and reproduces the
  experimental data, see Ref.{\protect \cite{bey96}}. Other lines are due
  to different nuclear densities, see text.}
\end{figure}

The cross section entering into Eq.~(\ref{eqn:liferes}) is given in
Fig.~\ref{fig:cross}. We restrict the two-body channels to the
dominant ones, i.e.  $^1S_0$ and $^3S_1- {^3D}_1 $. For the separable
ansatz we use the parameterization of Phillips~\cite{phi68}. The
parameters are taken from Ref.~\cite{bru77}, which lead to a good
overall description of the elastic and break-up cross sections as well
as the differential elastic cross section up to $E_{lab}=50$
MeV~\cite{bey96,sch83}.  To calculate the break-up cross section we
use the optical theorem.

The solid lines represent the isolated break-up cross section. As
shown previously it reproduces the experimental
data~\cite{bey96,sch83}.  The dashed lines show the break-up cross
section, $\sigma_{n,T}^*(E_{lab})$ for densities $n=0.1,\, 1,\, 3,\,
5,\, 7\, \times 10^{-3}$ fm$^{-3}$, respectively, as a function of the
laboratory energy $E_{lab}$. The Mott transition occurs at the density
$n\simeq 8\times 10^{-3}$ fm$^{-3}$.

\begin{figure}[p]
  \leavevmode
\centering
  \psfig{figure=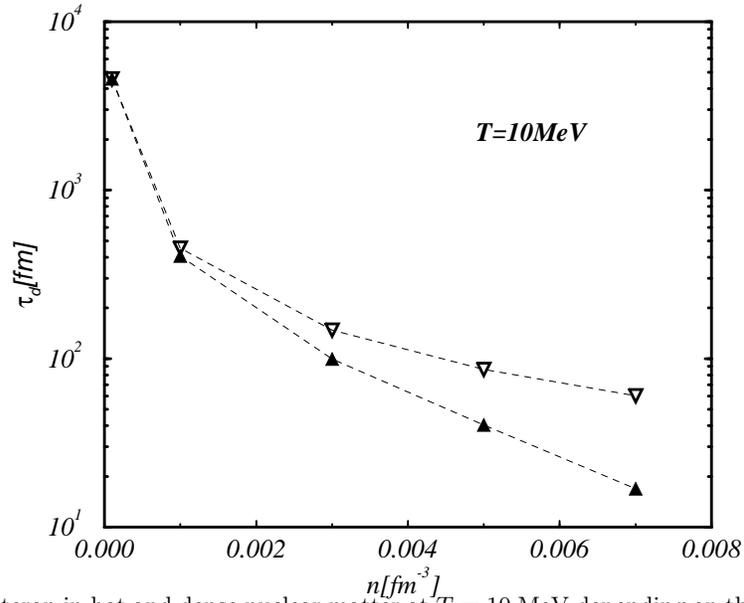,angle=270,width=0.6\textwidth}
\vspace*{1ex}
\caption{\label{fig:width} Width of the deuteron in hot and
  dense nuclear matter at $T=10$ MeV depending on the nuclear density
  and $P=0$ fm$^{-1}$.  The full triangles show the full calculation,
  the empty triangles show the one that uses the vaccuum cross section
  for the break-up reaction, free masses and deuteron binding energy.
  The deuteron binding energy is shown by the full diamonds. }
\end{figure}

The medium effects significantly modify the isolated break-up cross
section.  Two qualitative features are observed. First, the break-up
threshold is shifted towards lower scattering energies with increasing
density of the nuclear matter.  This kinematic effect is due to the
decrease of the deuteron binding energy with increasing density (see
Fig.~\ref{fig:width}).  Second, the cross section
increases considerably with increasing density. The maximum is
enhanced by one order of magnitude for the largest density value
considered.  For densities larger than the Mott densities the deuteron
disappears as a bound state.

Also, it is instructive to see how the medium dependent cross section
converges to the isolated one.  At $E_{lab}=100$ MeV the deviation of
the in-medium cross section from the free cross section in this model
is in the order of 10\%.  From inspection of Figure~\ref{fig:cross} we
conclude that the dominant changes in the cross section takes place at
rather moderate energies, i.e. where the impulse approximation fails,
and the Faddeev technique has to be used.

\begin{figure}[p]
  \leavevmode
\centering
   \psfig{figure=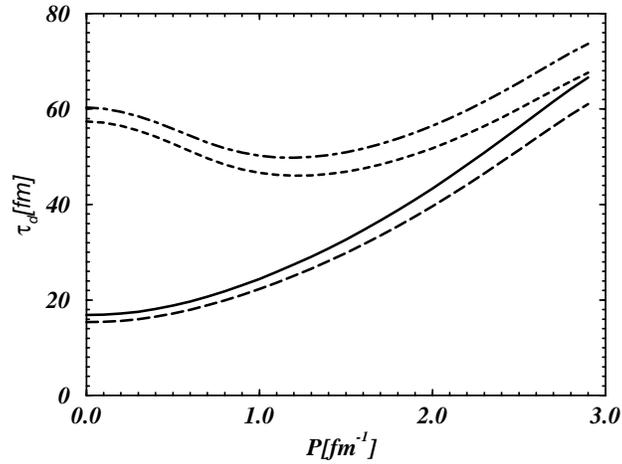,angle=270,width=0.6\textwidth}
\vspace*{1ex}
\caption{\label{fig:lifetime} Momentum dependent life-time $\tau_{dP}$ of
  the deuteron in hot and dense nuclear matter at $T=10$ MeV. Upper
  two curves are without medium modified cross section using the solid
  line of Fig.~\ref{fig:cross}, lower two lines with medium
  modifications for comparison. Solid and dashed-dotted lines uses
  Fermi distributions, the long dashed and short-dashed lines use
  classical distribution functions.}
\end{figure}

The resulting deuteron life-time evaluated with
Eq.~(\ref{eqn:liferes}) is shown in Fig.~\ref{fig:lifetime}. The
influence of the medium modification through the cross section is
substantial, in particular for small deuteron momenta. For higher
momenta the differences become smaller, as they
should, since the medium effects on the single particle properties
become smaller at higher momenta. Also, the difference between the
Maxwell and the Fermi distributions are shown that are comparably
small for the densities considered.

In Fig.~\ref{fig:width} the width of the deuteron at $P=0$ fm$^{-1}$
is shown as a function of nuclear density.
Respecting the medium effects in the cross section leads to a larger
width of the deuteron of almost a factor of two near the Mott density.
For comparison also the medium dependent deuteron binding energy is
shown in the same scale. 

\section{Summary and conclusions}

As already expected from earlier results on the NN-cross section,
we find that the deuteron break-up cross section ($Nd\rightarrow NNN$)
is also substantially modified at finite densities and temperatures
compared to the isolated one. The densities and the temperature
chosen are expected to be typical values for the final stage of heavy
ion collisions at intermediate energies. To reach this conclusion we
have extended the AGS-formalism to treat the effective three-nucleon
problem in an environment of hot and dense nuclear matter. This has
been achieved using the finite temperature Green-function method within the
Dyson-approach. The three-body problem is then formulated in the cluster
mean-field approximation, and the resulting AGS-equations are solved
numerically for a separable NN-potential.   For the isolated system the
experimental data are reproduced within a few per cent.

Within mean-field approximation the influence of the surrounding
matter leads 1) to a shift of the self energy of the nucleon and
deuterons and 2) to additional phase space factors due to Pauli
blocking.  This two effects are taken into account consistently as the
three-body equations are solved. 

The influence of the medium on the break-up cross section is
calculated in the center of mass frame for the three-particle system.
It shows three important features: 1) The cross section increases with
increasing densities, 2) the threshold energy shifts due to the
decreasing binding energy of the deuteron at increasing densities, and
3) the effect of the medium becomes smaller at higher scattering
energies.  The influence of the medium is rather strong. Near the
maximum and close to the Mott density the cross section increases
almost one order of magnitude compared to the isolated one.  We argue
that this modification is also important in a complete treatment of
the heavy ion reaction (as found for the NN-case~\cite{alm95}).

An important global quantity that governs the time scale of the
deuteron formation is the life-time (i.e. width) of the fluctuations
in the deuteron distributions in hot and dense nuclear matter. We have
calculated the life-time of the deuteron fluctuations using either
isolated or medium dependent cross section in the collision integral.
We find that the life-time strongly depends on the type of cross
section included in the evaluation. The difference between the use of
the isolated cross section versus the medium dependent one
amounts to almost a factor of three near the Mott density. This is in
support of the statement that the medium modifications of the break-up
cross section may lead to changes in the final outcome of the deuteron
rate in heavy-ion collisions.  Therefore the results presented here
may be considered an important input for transport equations that are
used to describe the dynamics in a heavy-ion collision as done, e.g.,
in Ref.~\cite{dan91}.

The formalism presented here is capable to be extended to effective
$n$-particle equations and therefore to treat the formation of higher
clusters than deuterons. In particular the formation of helium,
triton, or/and $\alpha$-particles is of special interest. In this
context, the inclusion of the total momentum dependence of the in-medium
cross section is necessary.

\section{Acknowledgment}

We are grateful to P. Schuck for discussions and valuable comments on the
manuscript as well as to V.G. Morosov and W. Sandhas for their interest.

\appendix

\section{Linear Response and the collision integral}
In order to derive the exact relation for the collision integral with
the full medium dependent cross section given in terms of properly
defined medium dependent three-particle transition operators, we
assume small fluctuations of the equilibrium distributions and utilize
the linear response theory to treat the nonequilibrium aspect of the
process.  

To do so consider first equilibrium. In this case the two-particle
distribution function may be decomposed in the following way,
reflecting the uncorrelated and the correlated part $g_2$,
\begin{equation}
f^0_2(12,1'2')=f^0_1(1)f^0_1(2)[\gd_{11'}\gd_{22'}-\gd_{12'}\gd_{21'}]
+ g^0_2(12,1'2'),
\label{eqn:twocor}
\end{equation}
where in cluster mean-field approximation,
\begin{equation}
g^0_2(12,1'2')
= \sum_{\nu P}\; 
 \left(\langle 12|\varphi_\nu\rangle 
\langle \varphi_\nu|1'2'\rangle
-\gd_{\nu P}(12)\gd_{\nu P}(1'2')\right)_{\rm ex}\; g(E_\nu),
\label{eqn:corr}
\end{equation}
and $\gd_{\nu P}(12) = \gd_{{\bf P},{\bf k}_1+{\bf k}_2}\gd_{({\bf
    k}_1-{\bf k}_2)/2,\nu}$, and $()_{\rm ex}$ denotes inclusion of
exchange terms.Using Eqs.~(\ref{eqn:twocor}) and (\ref{eqn:corr})
leads to
\begin{equation}
f^0_2(\nu,\nu')= g(E_\nu)\gd_{\nu,\nu'},
\end{equation}
i.e., $f^0_2(\nu,\nu')$ is diagonal in the indices $\mu$ that will be
used in the following.

In the framework of linear response the one- and two-particle
distributions may be characterized by the deviations from the
respective equilibrium distributions $f^0$ via small fluctuations
$\delta f$, viz.
\begin{eqnarray}
f_1(1;t) &= &f^0_1(1) + \gd f_1(1;t),\label{eqn:f1}\\ 
f_2(\nu,\nu';t) &= &f^0_2(\nu,\nu') + \gd f_2(\nu,\nu';t).
\label{eqn:f2} 
\end{eqnarray}
The relevant statistical operator~\cite{roe86} properly including the
one- and two-particle distributions is given by the generalized Gibbs
state,
\begin{equation}
\rho_{\rm rel} = \frac{1}{Z}\exp\left[-\gb(H-\mu N)
-\gb\sum_1 F_1(1,t)\gd n_1(1) 
-\gb\sum_{\nu'\nu} F_2(\nu,\nu',t)\gd n_2(\nu,\nu')  \right],
\label{eqn:gibbs}
\end{equation}
where the operators describing density fluctuations of the one- and two-particle
distributions appearing in Eq.~(\ref{eqn:gibbs}) are defined by
\begin{eqnarray}
\gd n_1(1) &=& a_1^\dagger a_1 
- \langle a_1^\dagger a_1 \rangle_0=n_1(1)-f_1^0(1),\\
\gd n_2(\nu,\nu') &= &b^\dagger_{\nu'} b_{\nu}- 
\langle b^\dagger_{\nu'} b_{\nu} \rangle_0
=n_2(\nu,\nu')-f_2^0(\nu,\nu').
\end{eqnarray}
The Lagrange parameters $F_1(t)$ and $F_2(t)$ of
Eq.~(\ref{eqn:gibbs}) are determined by the consistency relations
\begin{equation}
f_\gk(t) = {\rm Tr}\{\rho_{\rm rel}(t)\;n_\gk\},
\end{equation}
where $\gk$ collectively denotes the quantum numbers of the one- or
two-particle operators. Linearizing Eqs.~(\ref{eqn:f1}) and
(\ref{eqn:f2}), resp., with respect to  $F_\gk(t)$ leads to 
\begin{equation}
\gd f_\gk(t) = \sum_{\gk'}\;\gb F_{\gk'}(t)\; (n_\gk;n_{\gk'}),
\end{equation} 
where we have introduced the Kubo scalar product 
\begin{equation}
(A;B)=\frac{1}{\gb}\int\limits_0^\gb \;d\tau\;
{\rm Tr}\{\rho_0\;A(-i\tau)\;B\}.
\label{eqn:kubo}
\end{equation}
Furtheron we use the Laplace transform
\begin{equation}
\langle A(\eta); B\rangle
=\int_{-\infty}^0 dt \; e^{\eta t}\; (A(t);B).
\label{eqn:laplace}
\end{equation}

For the case of the deuteron density fluctuation, Eq.~(\ref{eqn:f2})
reads
\begin{equation}
\gd n_{dP} = b^\dagger_{dP}b_{dP}-g(E_{dP}),
\end{equation}
so that
\begin{equation}
\langle \gd n_{dP}\rangle^t = \gd f_{d}^{\rm R}(P,t).
\end{equation}
For small fluctuations $\gd f_{d}^{\rm R}(P,t)$ the response
parameters $F_{dP}(t)$ are small so that after linearizing we obtain
the explicit relation
\begin{equation}
\gd f_{d}^{\rm R}(P,t) = \sum_{P'}\;\gb F_{dP'}(t)\;
(\gd n_{dP'};\gd n_{dP}).
\label{eqn:linrel}
\end{equation}

The nonequilibrium statistical operator has the form (see
e.g.~\cite{mor95}),
\begin{equation}
\rho(t) = \rho_{rel}(t)
-\lim_{\eta\rightarrow 0^+}\int dt'\;e^{\eta(t'-t)}\;
U(t,t')\left\{i[H,\rho_{rel}(t')]
+\partial_{t'}\rho_{rel}(t')\right\}
U(t',t).
\label{eqn:rho}
\end{equation}
The occupation of the bound states tends to reach the equilibrium
value due to the reactions within the system. After linearization with
respect to $F_{dP}(t)$ and neglect of the explicit time dependence
(Markov limit) in the integral, viz. $\rho_{rel}(t')\simeq\rho_{rel}(t)$,
we obtain
\begin{equation}
\partial_t \,\gd f_{d}^{\rm R}(P,t)
=\langle i[H,n_{dP}]\rangle^t = I^{\rm R}_{dP}(t),
\label{eqn:resp}
\end{equation}
where we have introduced $I^{\rm R}_{dP}(t)\equiv I_{dN,NNN}(P,t)$ for
brevity.  Evaluating Eq.~(\ref{eqn:resp}) leads to
\begin{equation}
I^{\rm R}_{d}(P,t) = {\rm Tr}\{\rho(t)\;i[H,n_{dP}]\}
= \sum_{P'} \beta F_{nP'}(t)
\left[(n_{dP'};\dot n_{dP})-
\langle \dot n_{dP'};\dot n_{dP}\rangle\right].
\label{eqn:stoss}
\end{equation}
Note, that for homogeneous matter $[n_{dP'},n_{nP}]=0$, and than
$(n_{dP'};\dot n_{dP})=0$, which finally leads  to
Eq.~(\ref{eqn:fluk}) by using Eq.~(\ref{eqn:linrel}).

\section{Evaluation of the correlation function}

For the one-particle occupation $\dot n_{k}=i\,[H,n_k]$ we get
\begin{equation}
\dot n_{k}= i\sum_{121'2'}\langle 1|k\rangle\;
V_2(k2,1'2')\;
a^\dagger_{1} a^\dagger_{2} a_{2'} a_{1'} + h.c.
\label{eqn:ndot}
\end{equation}
In the three-particle space this leads to 
\begin{equation}
\dot n^{(\gc)}_{k} =
\sum_{1231'2'3'} \langle123 |\Lambda^{(\gc)}_k\bar V^{(\gc)}_3|1'2'3'\rangle\;
a^\dagger_{1} a^\dagger_{2} a^\dagger_{3} a_{3'}a_{2'} a_{1'} + h.c.,
\end{equation}
where we have introduced the third particle in the channel $\gc$, and
replaced $V_2(12,1'2')\;\langle 3|3'\rangle\;\rightarrow \bar
V^{(1)}_3(123,1'2'3')$ (for $\gc=1$), which has been defined in the
previous section. Further we use $V_2(12,1'2')=-V_2(21,1'2')$
etc. The projection operator $\Lambda^{(\gc)}_k$ is given by (e.g.
$\gc=1$)
\begin{equation}
\Lambda^{(1)}_k=|k23\rangle\langle k23|.
\end{equation}
The proper symmetrization of the final result will be treated
separately within the framework of the three-body formalism.

For $n^{(\gc)}_{dP}=b_{dP}b^\dagger_{dP}$ in channel $\gc$ the
time derivative is given by
\begin{equation}
\dot n^{(\gc)}_{dP} = -i\sum_{1231'2'3'}
\langle 123|\Lambda^{(\gc)}_{dP} \bar V^{(\gc)}_3|1'2'3'\rangle\;
a^\dagger_{1}a^\dagger_{2}a^\dagger_{3} a_{3'} a_{2'}a_{1'} + h.c.,
\label{eqn:ndeut}
\end{equation}
where we have introduced ($\gc=1$)
\begin{equation}
\Lambda^{(1)}_{dP}=2\;|1\varphi_{dP}\rangle\langle1\varphi_{dP}|.
\end{equation}
To keep the notation transparent, we introduce a particle-hole ($ph$)-basis,
i.e. $|123\rangle\otimes|\bar 1\bar 2\bar 3\rangle$ and
extent the potential in the following way, 
\begin{equation}
\Gamma^{(\gc)}_{6,a} = \Lambda^{(\gc)}_a\bar 
V^{(\gc)}_{3}\otimes {\sf I}_3,\qquad
\tilde \Gamma^{(\gc)}_{6,a} = {\sf I}_3\otimes
\bar V^{(\gc)}_{3}\Lambda_a^{(\gc)}, 
\end{equation}
with $a=\{k,dP\}$.

\begin{figure}[p]
  \leavevmode
\centering
   \psfig{figure=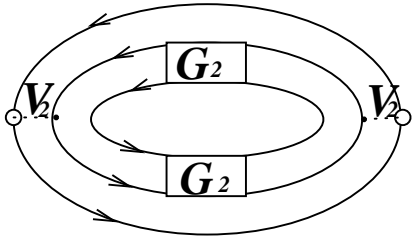,angle=270,width=0.3\textwidth}
\vspace*{1ex}
\caption{\label{fig:G6Born} Pictorial demonstration of the 
  Born approximation for $G_6(\gO_\mu)$. The dots indicate the
  Potential $V_2$. Exchange and rearrangement channels have to be
  added for a full treatment.}
\end{figure}

\begin{figure}[p]
  \leavevmode
\centering
  \psfig{figure=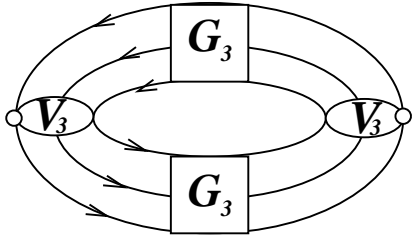,angle=270,width=0.3\textwidth}
\vspace*{1ex}
\caption{\label{fig:G6AGS} Approximation of $G_6$ with full treatment
  of the intermediate three-particle Green function.}
\end{figure}

Due to the operators occuring in Eq.~(\ref{eqn:ndeut}) the correlation
function $ \langle \dot n;\dot n\rangle$ is related to the 12-point
Green function $G_6(\gO_\mu)$ via Eq.~(\ref{eqn:Gcorr}). Using
Eqs.~(\ref{eqn:ndot}) and (\ref{eqn:ndeut}) the Green function
$G_{\dot n_{k} \dot n_{k}} (\gO_\mu)$ reads
\begin{equation}
G^{(\gc)}_{\dot n_{a} \dot n_{a}} (\gO_\mu) =
{\rm Tr}\left\{ \Gamma_{6,a}^{(\gc)}\; G_6 (\gO_\mu)\Pi_{ph} \; 
\tilde \Gamma_{6,a}^{(\gc)} \right\}
+ (\gO_\nu\leftrightarrow -\gO_\nu),
\label{eqn:GR1}
\end{equation}
and $\Pi_{ph}$ exchanges all particle with hole indices.  The
contribution with $\gO_\nu\leftrightarrow -\gO_\nu$ is due to the
($h.c.$) that appear in Eq.~(\ref{eqn:ndeut}). If only three-particle
correlations are considered the  full
12-point Green function is given by
\begin{equation}
G_6(\gO_\mu)=\frac{1}{-i\gb}\sum_\gl G_3(\gO_\mu+z_\gl)\otimes
G_3 (z_\gl),
\end{equation}
where explicit Matsubara summation has been introduced. Then 
summation over the block indices can be performed so that the resulting
traces are in three-particle space only. Eq.~(\ref{eqn:GR1}) 
simplifies to
\begin{eqnarray}
G^{(\gc)}_{\dot n_{a} \dot n_{a}} (\gO_\mu) 
&= &\frac{1}{-i\gb}\sum_\gl
{\rm Tr}\left\{\;\Lambda^{(\gc)}_a\bar V^{(\gc)}_{3}\;
G_3(\gO_\mu+z_\gl)
\;\bar V^{(\gc)}_{3}\Lambda^{(\gc)}_a 
\;G_3(z_\gl)\right\}
\label{eqn:G6totN}\\
&&\qquad+ (\gO_\nu\leftrightarrow -\gO_\nu).\nonumber
\end{eqnarray}
In Eq.~(\ref{eqn:G6totN}) the term
$\Lambda^{(\gc)}_a G_3(z_\gl)\Lambda^{(\gc)}_a$ appears.
For $a=k$ and e.g. $\gc=1$ this is given by 
\begin{equation}
\langle 123|\Lambda^{(1)}_k G_3\Lambda^{(1)}_k|1'2'3'\rangle
=G_3(k23,k2'3').
\end{equation}
For $a=dP$ we first consider $G_3\Lambda^{(\gc)}_{dP}$ using
Eq.~(\ref{eqn:G3Ggam}), i.e.
\begin{equation}
G_3\Lambda^{(\gc)}_{dP} 
 = G_3^{(\gc)}\Lambda^{(\gc)}_{dP}
+G_3 \bar V_3^{(\gc)} G^{(\gc)}_3\Lambda^{(\gc)}_{dP}.
\label{eqn:G3GgamL}
\end{equation}
To evaluate this expression we introduce the spectral decomposition of
$G_3^{(\gc)}$,
\begin{equation} 
G^{(\gc)}_3(z_\gl)=  
\sum_n \frac{|\phi^{(\gc)}_n\rangle\langle\phi^{(\gc)}_n|}
          {(z_\gl -\gve_\gc) -E^{(\gc)}_{n}}
\;(1-f_\gc+g(E^{(\gc)}_{n})).
\label{eqn:specB}
\end{equation}
The wave function given in Eq.~(\ref{eqn:specB}) is a direct product
of noninteracting wave functions of the pair and the odd particle,
e.g. $|\phi^{(1)}_n\rangle = |1\varphi_n\rangle$. Through
orthogonality, only the bound state part with momentum $P$
contributes.  Similar arguments hold for the projection from the left
side. We will denote this cluster Green function
by $G_{3,dP}$. Note that through Eq.~(\ref{eqn:G3GgamL})
etc. the full scattering solution is taken into account.
Eq.~(\ref{eqn:G6totN}) then reads
\begin{eqnarray}
G^{(\gc)}_{\dot n_{dP} \dot n_{dP}} (\gO_\mu) 
&= &\frac{1}{-i\gb}\sum_\gl
{\rm Tr}\left\{\;\bar V^{(\gc)}_{3}\;
G_{3,0}(\gO_\mu+z_\gl)
\;\bar V^{(\gc)}_{3}\;G_{3,dP}(z_\gl)\right\}
\label{eqn:G6neu}\\
&&\qquad+ (\gO_\nu\leftrightarrow -\gO_\nu).\nonumber
\end{eqnarray}
Since we are only interested in the break-up reaction we have
introduced an extended notation also for the second Green function
appearing in Eq.~(\ref{eqn:G6neu}), $G_{3,0}$ denotes the full Green
function that describes the break-up situation. The related diagram is
given in Fig.~\ref{fig:G6AGS}. The Born approximation is given by
Fig.~\ref{fig:G6Born}.

It is important to note, that through Eq.~(\ref{eqn:Gcorr}) we are
only interested in the discontinuity of $G_{\dot n_k \dot n_{k'}}(z)$
on the real axis, which leads to energy conservation. Therefore we
eventually need to consider only the on-energy-shell limit.  It is now
possible to introduce the break-up operator $U_{0\gc}$ given in the
Section III. Using the on-energy-shell requirements, 
\begin{eqnarray}
U_{0\gc} G^{\gc}_3 &=& \bar V^{\gc}_3G_{3,dP},\\
U_{\gc 0} G^{(0)}_3 &=& \bar V^{\gc}_3G_{3,0},
\end{eqnarray}
these equations lead directly to Eq.~(\ref{eqn:G6full}) and
Eq.~(\ref{eqn:G6fine}). 
For small deviations from equilibrium we may consider, e.g., the
contribution of the loss term (first term in brackets of
Eq.~(\ref{eqn:G6fine})) to determine the
scale of the life time of the fluctuations. The resulting expression
for the deuteron distribution in hot and dense nuclear
matter is then given by
\begin{equation}
\tau^{-1}_{dP} =
\langle \gd \dot n_{dP};\gd\dot
n_{dP}\rangle \;g^{-1}(E_{dP}).
\end{equation}
Here we have used that in ladder $t$-matrix approximation
\begin{equation}
(n_{dP};n_{dP}) = g(E_{dP}).
\end{equation}
This completes the deviation of Eq.~(\ref{eqn:corrfunc}).

\newpage

\end{document}